\documentclass[preprint,aps,12pt,showpacs,nofootinbib,tightenlines]{revtex4}
\usepackage{amsmath}
\usepackage{amssymb}
\usepackage{epsfig}
\usepackage{graphicx}
\begin{document}
\def\pslash{\rlap{\hspace{0.02cm}/}{p}}
\def\eslash{\rlap{\hspace{0.02cm}/}{e}}
\def\Eslash{\rlap{\hspace{0.02cm}/}{E}}
\let\jnfont=\rm
\def\NPB#1,{{\jnfont Nucl.\ Phys.\ B }{\bf #1},}
\def\PLB#1,{{\jnfont Phys.\ Lett.\ B }{\bf #1},}
\def\EPJC#1,{{\jnfont Eur.\ Phys.\ Jour.\ C }{\bf #1},}
\def\PRD#1,{{\jnfont Phys.\ Rev.\ D }{\bf #1},}
\def\PRL#1,{{\jnfont Phys.\ Rev.\ Lett.\ }{\bf #1},}
\def\MPLA#1,{{\jnfont Mod.\ Phys.\ Lett.\ A }{\bf #1},}
\def\IJMPA#1,{{\jnfont Int.\ J.\ Mod.\ Phys.\ A }{\bf #1},}
\def\JPG#1,{{\jnfont J.\ Phys.\ G }{\bf #1},}
\def\CTP#1,{{\jnfont Commun.\ Theor.\ Phys.\ }{\bf #1},}
\def\CPL#1,{{\jnfont Chin.\ Phys.\ Lett.\ }{\bf #1},}
\def\CHIC#1,{{\jnfont Chin.\ Phys.\ C.\ }{\bf #1},}
\def\JHEP#1,{{\jnfont JHEP \ }{\bf #1},}
\def\NPPS#1,{{\jnfont Nucl.\ Phys.\ Proc.\ Suppl.\ }{\bf #1},}
\def\CPC#1,{{\jnfont Computl.\ Phys.\ Commun.\ }{\bf #1},}
\def\APPB#1,{{\jnfont Acta\ Phys.\ Polon.\ B }{\bf #1},}

\def\EPL#1,{{\jnfont Europhys.\ Lett. }{\bf #1},}

\def\btt#1{{tt$\backslash$#1}}
\def\BibTeX{\rm B{\sc ib}\TeX}
\title{ Pair production of charged  top-pions
in the $\gamma\gamma$ collisions at the ILC}
\author{Jinzhong Han}
\author{Xuelei Wang}
\affiliation{\small
College of Physics and Information Engineering, Henan Normal
University, Xinxiang, Henan 453007, P.R.China}

\begin{abstract}

 The top-color assisted technicolor (TC2) mode predicts the existence of
a pair of charged top-pions $\pi^{\pm}_t$. In this paper, we study
the production of the charged top-pions pair $\pi^{\pm}_t$ at next
generation $\gamma\gamma$ colliders. The results show that the
production rates can reach the level of $10^2$ fb with reasonable
parameter space. With a large number of events and the clean
background, the charged top-pion should be observable at the
$\gamma\gamma$ colliders. Therefore, our studies can help us to
search for charged top-pion, and furthermore, to test the TC2 model.

\end{abstract}

\maketitle

\section{ Introduction}

It is widely believed that the hadron colliders, such as Tevatron
and LHC, can directly probe possible new physics beyond the standard
model (SM) up to a few TeV, while the International Linear Collider
(ILC) can produce new physics signal events more easily resolved
from backgrounds \cite{ILC}. An  $e^+e^-$  ILC can also be designed
to operate as a $\gamma\gamma$ collider. This is achieved by using
Compton backscattered photons in the scattering of intense laser
photons on the initial $e^+e^-$ beams.
 In this
case, the energy and luminosity of the photon beam would be the same
order of magnitude of the parent electron beam and the set of final
states are much richer than that in the $e^+e^-$ mode. Therefore,
the $\gamma\gamma$ collider will provide us a good chance to pursuit
new physics particles.


As a possible solution to avoid the shortcomings of the triviality
and unnaturalness arising from the elementary Higgs field, the
technicolor theory was proposed and so far remains a popular
candidate for new physics beyond SM \cite{TC}. Among the various TC
models, the topcolor-assisted technicolor (TC2) model is a more
realistic one \cite{TC2,EPS}, which is consistent with the current
experiment. In the TC2 model, the topcolor interaction makes small
contribution to the electroweak symmetry breaking (EWSB), and gives
rise to the main part of the top quark mass $(1-\varepsilon)m_t$
with a model dependant parameter $0.03\leq\varepsilon\leq0.1$
\cite{EPS}. One of the most general predictions of the TC2 model is
the existence of three Pseudo-Goldstone bosons, so called top-pions
($\pi^{\pm}_t,\pi^{0}_t$) and an isospin singlet boson called
top-higgs ($h^{0}_t$), which masses are in the range of hundreds of
GeV. Thus, studying the possible signatures of these typical
particles in the high energy experiments would provide crucial
information for EWSB and fermion flavor physics. Furthermore, the
discovery of these new particles can be regarded as direct evidence
to test the TC2 model. The TC2 model is expected to give new
significant signatures in future high energy colliders and studied
in references \cite{TC-WYC1,TC-WYC2,TC-WYC3}, due to the new
particles which are predicted by this model. Because the SM predicts
the existence of one neutral Higgs  boson,
so any  observation  of charged Higgs
particles will
mean the signal of a new physics. Therefore, probing of charged
top-pions is  more  important  to test the  TC2  model.  At  the
high-energy $e^+e^-$ linear colliders, the main signal charged
top-pion production processes are $e^+e^-(\gamma\gamma)\rightarrow
b{\bar c}\pi^{+}_t$, $e^+e^-(\gamma\gamma)\rightarrow t{\bar
b}\pi^{-}_t$, and $e^+e^-\rightarrow W^{+}_t\pi^{-}_t$, which have
been systematically studied in ref. \cite{signal-pi}.

So far, the reaction $e^+e^-\rightarrow hh$  have been studied in
the SM \cite{SM-hh}, and the similar processes for the neutral or
charged scalar have also been investigated via $e^+e^-$ or
$\gamma\gamma$ model in the minimal supersymmetric standard model
(MSSM) \cite{MSSM-hh}, the two-Higgss doublet model (2HDM)
\cite{2HDM-hh}, the littlest Higgs (LH) model \cite{LH-hh}, as well
as the left-right twin Higgs (LRTH) model
\cite{LRTH-hh-1,LRTH-hh-2}. In the TC2 model the pair production of
the charged scalar $e^+e^-\rightarrow \pi^+_t\pi^-_t$ has also been
probed in Ref. \cite{ee-pipi}. Complementing Ref. \cite{ee-pipi}, in
this paper we will study the production of the pair charged
top-pions via process $\gamma\gamma\rightarrow \pi^+_t\pi^-_t$ at
$\gamma\gamma$ colliders within the TC2 model.

\indent The remainder parts of this paper are organized as follows.
In Sec II, we shall present the calculations of the production cross
section of the process $\gamma\gamma\rightarrow \pi^{+}_t\pi^{-}_t$.
Numerical results on the cross section and  concluding remarks will
appear in Sec. III.
\section{Calculations of Production cross section }
 The Feynman diagrams for the process $\gamma\gamma\rightarrow \pi^{+}_t\pi^{-}_t$
 are given in Fig. 1. And the relevant Feynman rules for the couplings
 $A^\mu\pi^{+}_t\pi^{-}_t$ and $A^\mu A^\nu\pi^{+}_t\pi^{-}_t$
  in the TC2 model can be found in Ref.
\cite{coupling}.

With the relevant couplings, the invariant production amplitudes of
the process $e^-(p_1)e^+(p_2)\rightarrow
\pi^{+}_t(p_3)\pi^{-}_t(p_4)$ can be written as follows:

\begin{figure}[htb]
\scalebox{0.35}{\epsfig{file=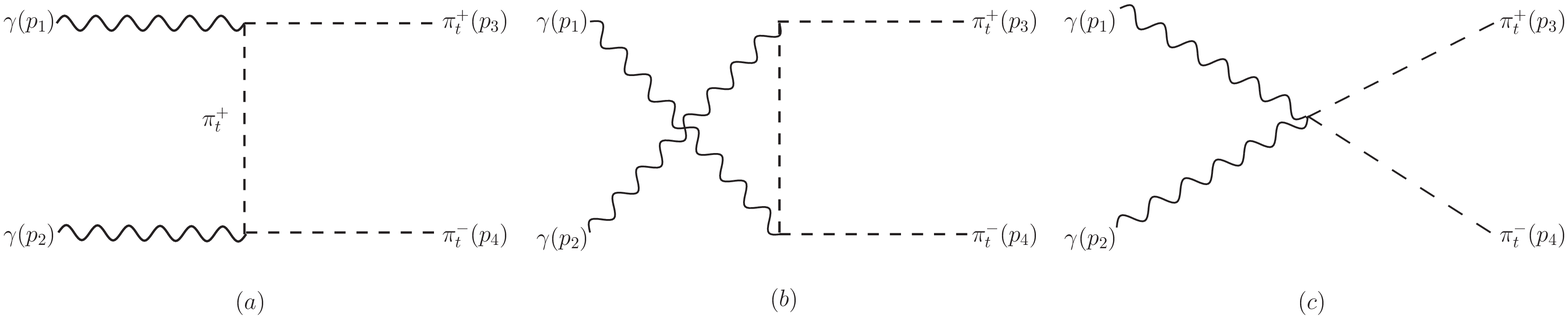}} \caption{Feynman diagrams of
the processes  $e^+e^-\rightarrow W^{\pm}{\pi}^{\pm}$ in the TC2
model.}
\end{figure}

\begin{eqnarray}
M_a=\frac{ie^2}{(p_3-p_1)^2-m_{\pi}^2}(p_1-2p_3)^\mu\epsilon_\mu(p_1)(2p_4-p_2)^\nu\epsilon_\nu(p_2)\\
M_b=\frac{ie^2}{(p_3-p_2)^2-m_{\pi}^2}(p_2-2p_3)^\mu\epsilon_\mu(p_2)(2p_4-p_1)^\nu\epsilon_\nu(p_1)\\
M_c=2ie^2g^{\mu\nu}\epsilon_\mu(p_1)\epsilon_\nu(p_2)~~~~~~~~~~~~~~~~~~~~~~~~~~~~~~~~~~~~~~~~~
\end{eqnarray}
Here, $\epsilon_\mu(p_1)$ and $\epsilon_\nu(p_2)$ are the
polarization vector of the photon, $m_\pi{}$ denotes the mass of the
charged top pions.

With the above amplitudes $M_a,~M_b$ and  $M_c$, we can directly
obtain the production cross section $\hat{\sigma}(\hat{s})$ for the
subprocess $\gamma\gamma\rightarrow \pi^{+}_t\pi^{-}_t$ and the
total cross sections at the $e^+e^-$ linear collider can be obtained
by folding $\hat{\sigma}(\hat{s})$ with the photon distribution
function $F(x)$ which is given in Ref. \cite{distribution},
\begin{eqnarray}
\sigma_{tot}(s)=\int^{x_{max}}_{{E_0}/{\sqrt{s}}}dz\frac{{\rm
d}{\cal L}_{\gamma \gamma}} {{ dz} } \hat{\sigma}({\gamma \gamma \to
\pi^{+}_t\pi^{-}_t}, {\rm at}~ {\hat{s}=z^2s})
\end{eqnarray}
where $E_0=2m_{\pi}$, $s$ is the squared c. m. of $e^+e^-$
collision,
\begin{eqnarray}
\frac{{\rm d}{\cal L}_{\gamma \gamma}} {{ dz}
}&=&2z\int^{x_{max}}_{{z^2}/{x_{max}}} \frac{{ dx}} {{x}
}F_{\gamma/e}(x)F_{\gamma/e}(\frac{z^2}{x}), \label{definition}
\end{eqnarray}
For the initial unpolarized electrons and laser photon beams, the
energy spectrum of the backscattered photon is given by
\begin{eqnarray}
\displaystyle F_{\gamma/e}=\frac{1}{D(\xi)}\left[1-x+\frac{1}{1-x}
-\frac{4x}{\xi(1-x)}+\frac{4x^2}{\xi^2(1-x)^2}\right],
\end{eqnarray}
with
\begin{eqnarray}
\displaystyle D(\xi)=\left(1-\frac{4}{\xi}-\frac{8}{\xi^2}\right)
\ln(1+\xi)+\frac{1}{2}+\frac{8}{\xi}-\frac{1}{2(1+\xi)^2},
\end{eqnarray}
The definitions of parameters $\xi$, $D(\xi)$ and $x_{max}$ can be
found in \cite{distribution}. In our numerical calculation, we
choose $\xi=4.8$, $D(\xi)=1.83$ and $x_{max}=0.83$.

\section{numerical results and conclusions }

To get the numerical results of the cross section, we should also
fix  the parameter in the SM as $\alpha_e=1/128.8$
\cite{parameters}. In addition, the top-pions mass $m_{\pi}$ is also
involved in the production amplitudes. The TC2 model loosely
predicts top-pions to lie in the mass range of about $100-300$ GeV.
Top-pions this light are disfavored by the data for $R_b$
\cite{R-b}, but the contribution of new ETC gauge bosons can help to
relax the constraint on the top-pions mass \cite{mpi}. In this work,
we take $m_{\pi}$ as a free parameter and expand the mass range to
$150-350$ GeV to estimate the total cross section of
$\pi^{+}_t\pi^{-}_t$ associated production at the ILC.

\begin{figure}[htbp]
\includegraphics[width=3.0in,height=3.0in]{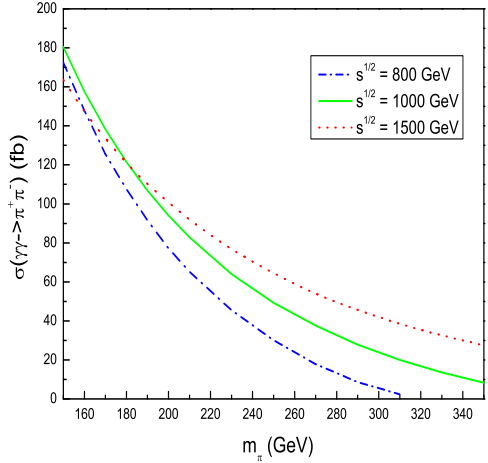}
\vspace{-0.3cm} \caption{The production cross section of the
processes $\gamma\gamma\rightarrow \pi^{+}_t\pi^{-}_t$ in the TC2
model as a functions of $m_{\pi}$.}
\end{figure}

In Fig. 2, we plot the cross section $\sigma$ of the process
$\gamma\gamma\rightarrow \pi^{+}_t\pi^{-}_t$ as a function of the
mass parameter $m_{\pi}$ with the three values of the center of mass
energy. The plot shows that the cross section $\sigma$ decreases
with $m_{\pi}$ increasing, this is because the phase space is
depressed strongly by large $m_{\pi}$.  In general, the production
rate is at the level of $10^{1}\sim 10^2$ fb in a large part of the
allowed parameter space. For $\sqrt{s}=1000$ GeV, $\sqrt{s}=1500$
GeV, and for $150 ~\rm{GeV}\leq M_{\pi}\leq 350 ~\rm{GeV}$, the
value of $\sigma$ is in the range of $8.3\sim180.5$ fb and
$27.5\sim163.7$ fb, respectively. According to the ILC Reference
Design Report \cite{ILC}, the ILC is determined to run with
$\sqrt{s}=500$ GeV (upgradeable to $1000$ GeV) and the total
luminosity required is $\mathcal {L}=500\ {\rm fb}^{-1}$ with the
first four-year operation and $\mathcal {L}=1000\ {\rm fb}^{-1}$
during the first phase of operation with $\sqrt{s}=500$ GeV. If we
assume the intergrated luminosity $\mathcal {L}=500~{\rm fb}^{-1}$,
 there will be  up to several hundreds of
$\pi^{+}_t\pi^{-}_t$ events to be generated per year at the ILC
experiments.

\begin{figure}[htbp]
\includegraphics[width=3.0in,height=3.0in]{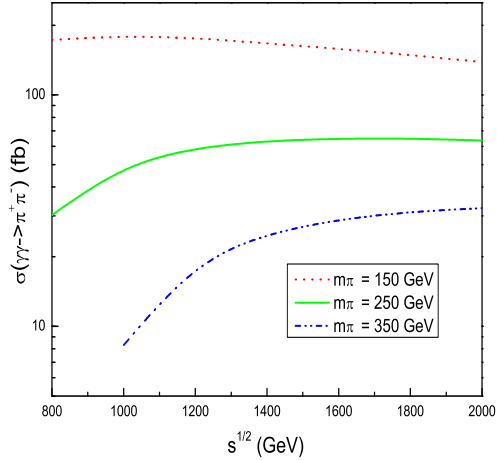}
\vspace{-0.3cm}
 \caption{The production cross section of the processes
$\gamma\gamma\rightarrow \pi^{+}_t\pi^{-}_t$ in the TC2 model as a
functions of  $\sqrt{s}$.}
\end{figure}

To see the effect of the c.m. energy on the production cross section
, in Fig.3 we plot the cross section $\sigma$ as the function of
$\sqrt{s}$ with  $m_{\pi}=150,~250$ and $350$ GeV, respectively. Due
to the contributions to the cross section come mainly from the
t-channel and u-channel, thus the large $\sqrt{s}$ can enhance the
cross section significantly and the values of the cross section can
reach 180 fb maximally in the reasonable parameter space.

Considering the subsequent main decay of $\pi^{+}_t\rightarrow
t\bar{b}$ \cite{pi-tb} and $t\rightarrow W^{+}b\rightarrow l^{+}\nu
b$, the possible signal for $\pi^{+}_t\pi^{-}_t$ production at the
ILC is four jets $b\bar{b}b\bar{b}$ + two leptons + missing energy
$\Eslash_T$. The production rate of the $t\bar{t}b\bar{b}$ final
state can be easily estimated
$\sigma^{s}\approx\sigma\times[Br(\pi^{+}_t\rightarrow
t\bar{b})\times Br(\pi^{-}_t\rightarrow \bar{t}b)]$. Using the value
of the branching ratio $B_{r}(\pi^{+}_t\rightarrow t\bar{b})$
\cite{pi-tb} which is affected by the model-dependent parameter
$\varepsilon$, in Fig. 4, we show the numerical results for the
production rate of the $t\bar{t}b\bar{b}$. From Fig. 4, one can see
that the production rate can reach 50 fb with reasonable values of
the free parameters of the TC2 model.The cross section of the
irreducible $t\bar{t}b\bar{b}$ background  for $\sqrt{s}$= 800 GeV
 has been estimated in reference \cite{LRTH-hh-1} and found to be 5.5 fb.
  Thus, it may be possible to extract the signals from the
backgrounds in the reasonable parameter space of the TC2 model by
considering kinematic distribution of the signal. Certainly,
detailed confirmation of the observability of the signals generated
by the process $\gamma\gamma\rightarrow \pi^{+}_t\pi^{-}_t$ would
depend on Monte Carlo simulations of the signals and backgrounds,
which is beyond the scope of this paper.

\begin{figure}[htbp]
\includegraphics[width=3.0in,height=3.0in]{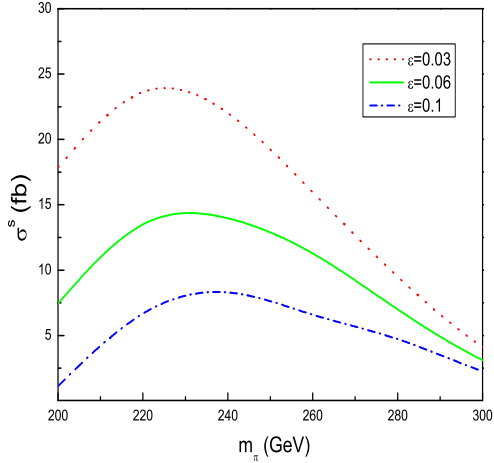}
\vspace{-0.3cm} \caption{The production rate of the
$t\bar{t}b{\bar{b}}$ finial state of the process
$\gamma\gamma\rightarrow \pi^{+}_t\pi^{-}_t$ as a function of the
parameter $m_{\pi}$ for $\sqrt{s}$ = 1000 GeV with
$\varepsilon=0.03$, $0.06$ and $0.1$, respectively.}
\end{figure}

For the light charged top-pions, the branching ratio of
$\pi^{+}_t\rightarrow c\bar{b}$ can be comparative to that of
$\pi^{+}_t\rightarrow t\bar{b}$. In this case, $\pi^{+}_t\rightarrow
c\bar{b}$ is also an important mode which induces the signals
$c\bar{b}\bar{c}b$. Although $\pi^{+}_t\rightarrow c\bar{b}$ is a
flavor-changing decay mode, $c\bar{b}\bar{c}b$ production is not the
flavor-changing process. Therefore, the SM background can not be
ignored. The major irreducible background should come from
$e^+e^-\rightarrow ZZ\rightarrow c\bar{c}b\bar{b}$. The mistagging
of b-quark and s-quark will make the $e^+e^-\rightarrow W^{+}W^{-}$
become important which significantly enhances the background. So,
the efficient b-tagging and mass reconstruction of the charged
top-pion are very necessary to reduce the background.

In summary, in the TC2 model we evaluate the signal pair production
of charged top-pions  via process $\gamma\gamma\rightarrow
\pi^{+}_t\pi^{-}_t$ at the ILC. It is found that TC2 model can make
a significant contribution to this processes. For the process
$\gamma\gamma\rightarrow \pi^{+}_t\pi^{-}_t$, can reach $10^2$ fb in
optimal case. Considering the main decay mode of top-pions, we find
that with such cross section, it is possible to detect the signal of
the charged top-pions experimentally in the most of the parameter
space at the future linear linear colliders operating in
$\gamma\gamma$ model at the TeV  energy scale. Even if we can not
observe the signals in future ILC experiments, at least, we can
obtain the bounds on the free parameters of the TC2 model.
\section{Acknowledgments}
 This work is supported by the National Natural Science
Foundation of China under Grant No. 11075045, and by Specialized
Research Fund for the Doctoral Program of Higher Education under
Grant No.20094104110001.

\end{document}